\documentclass[conference]{IEEEtran}
\usepackage{cite}
\usepackage{amsmath,amssymb,amsfonts}
\usepackage{algorithm} 
\usepackage{algorithmic}

\usepackage{graphicx}
\usepackage{textcomp}
\usepackage{xcolor}
\ifCLASSOPTIONcompsoc
\usepackage[caption=false,font=normalsize,labelfon
t=sf,textfont=sf]{subfig}
\else
\usepackage[caption=false,font=footnotesize]{subfi
g}
\fi
\def\BibTeX{{\rm B\kern-.05em{\sc i\kern-.025em b}\kern-.08em
    T\kern-.1667em\lower.7ex\hbox{E}\kern-.125emX}}
\begin{document}

\title{Message Scheduling for Performant, Many-Core Belief Propagation}

\author{\IEEEauthorblockN{Mark Van der Merwe, Vinu Joseph, Ganesh Gopalakrishnan}
\IEEEauthorblockA{\textit{School of Computing} \\
\textit{University of Utah}\\
Salt Lake City, USA \\
mark.vandermerwe@utah.edu,\{vinu,ganesh\}@cs.utah.edu}
}

\maketitle

\begin{abstract}
Belief Propagation (BP) is a message-passing algorithm for approximate inference over Probabilistic Graphical Models (PGMs), finding many applications such as computer vision, error-correcting codes, and protein-folding. While general, the convergence and speed of the algorithm has limited its practical use on difficult inference problems. As an algorithm that is highly amenable to parallelization, many-core Graphical Processing Units (GPUs) could significantly improve BP performance. Improving BP through many-core systems is non-trivial: the scheduling of messages in the algorithm strongly affects performance. We present a study of message scheduling for BP on GPUs. We demonstrate that BP exhibits a tradeoff between speed and convergence based on parallelism and show that existing message schedulings are not able to utilize this tradeoff. To this end, we present a novel randomized message scheduling approach, Randomized BP (RnBP), which outperforms existing methods on the GPU.
\end{abstract}


\section{Introduction}

Probabilistic Graphical Models (PGMs) are powerful, general machine learning models that encode distributions over random variables. PGM Inference, in which we seek to compute some probabilistic beliefs within the system modeled by the PGM, is in general an intractable problem, leading to dependence on approximate algorithms. Belief Propagation (BP) is a widely employed approximate inference algorithms for PGMs \cite{yedidia2001generalized}. BP has been successfully utilized in many areas, including computer vision \cite{felzenszwalb2006efficient}, error-correcting codes \cite{romero2012sequential}, and protein-folding \cite{yanover2003approximate}.

BP is a message-passing algorithm, in which messages are passed along edges of the PGM graph. While BP is exact on tree PGMs, it is approximate on general graphs containing loops, where iterative updates are applied until convergence. Like others \cite{elidan2006residual}, we break the performance of BP into two properties: convergence (for how many input graphs does it reach a convergent state) and speed (how long does it take to reach the convergent state). While BP has been shown to perform well on many graphs containing loops, there is no guarantee of convergence in most cases, and graphs of varying structure and parameterization can prevent BP from converging or can have slow speed for convergence \cite{murphy1999loopy}.

General-Purpose GPU computing has begun recently exploring many-core parallelism for graph-based problems \cite{wang2017gunrock}. This, combined with the inherent parallelism available between message updates, suggests that many-core parallelism can be effectively applied to BP to yield good performance on the GPU (that is, good convergence and speed). In order to ensure good performance, one must be careful in the implementation such as to avoid the convergence and speed pitfalls inherently present in Belief Propagation.

In existing BP literature, there has been much interest in exploring the use of message schedulings for improving BP performance. The naive scheduling is known as Synchronous or Loopy BP (LBP), where all messages are updated in parallel \cite{murphy1999loopy}. Asynchronous approaches, where some amount of sequentiality is enforced during the message updates, for example via subgraph updates \cite{wainwright2003tree} or greedy message selection \cite{elidan2006residual,pmlr-v5-gonzalez09a}, have been shown both empirically and theoretically to outperform LBP in single-core environments. The general intuition is that enforcing sequentialism in the scheduling encourages more direct propagation of information, thus converging faster. The contrast between LBP and Asynchronous BP introduces a parallelism vs. efficiency spectrum (also found in other graph problems such as SSSP \cite{davidson2014work}). LBP exposes high levels of parallelism but is work-inefficient. Asynchronous BP is efficient and convergent but exposes little parallelism. We hypothesize that there exists a tradeoff between the parallelism and sequentialism in Belief Propagation, and that GPUs can effectively harness that tradeoff to yield performant BP.

We start by presenting many-core frontier-based implementations for two greedy asynchronous message schedulings, Residual Belief Propagation \cite{elidan2006residual} and Residual Splash \cite{pmlr-v5-gonzalez09a}. We then benchmark the performance, varying parallelism to explore how parallelism affects the performance of BP. As expected, we find that as parallelism is increased, we see less convergence but obtain faster speed. As parallelism is decreased, we see more convergence but lower speeds. This is encouraging, as it means we can still get convergence boosts while exploiting parallelism, but we also see that existing approaches incur significant overheads, and performance is heavily tied to the choice of parallelism. To overcome these drawbacks, we propose a new message scheduling, called Randomized Belief Propagation (RnBP) which uses low-overhead, randomized scheduling, and outperforms existing approaches.

To summarize, our contributions are:
\begin{itemize}
    \item Many-core, frontier-based implementations of two greedy asynchronous message schedulings, Residual Belief Propagation \cite{elidan2006residual} and Residual Splash \cite{pmlr-v5-gonzalez09a}.
    
    \item Demonstration of tradeoff between parallelism and sequentialism in terms of speed/convergence of BP.
    
    \item Demonstration that overheads prevent existing asynchronous message scheduling approaches from scaling to the GPU. 
    
    \item A novel message-scheduling, Randomized BP (RnBP), that utilizes randomization to effectively handle the tradeoff BP demonstrates. We demonstrate speedups on Ising grid \cite{elidan2006residual} and protein-folding \cite{yanover2003approximate} datasets.
\end{itemize}

\section{Background} \label{sec:background}

\subsection{Belief Propagation}

We focus our attention on the Sum-Product Belief Propagation algorithm over discrete pairwise Markov Random Fields (MRFs), though we expect the results to generalize to other variants of BP. Suppose we have the set of discrete random variables $X=\{X_1,...,X_n\}$, each taking on a value $X_i\in A_i$, where $A_i$ is a finite set. An MRF is an undirected graph $G=(V,E)$. Each vertex $i\in V$ represents a discrete random variable $X_i$. $\{\psi_i:A_i\to \mathbb{R}^+ | i\in V\}$ is set of unary potential functions for each random variable. Each edge $(i,j)\in E$ represents the probabilistic relationship between two variables. $\{\psi_{i,j}:A_i\times A_j\to \mathbb{R}^+ | (i,j)\in E\}$ is the set of binary potential functions for each edge. An MRF yields the following joint distribution over $X$:
\begin{equation}
    P(x_1,...,x_N) \propto \prod_{i \in V} \psi_i(x_i) \prod_{\{i,j\}\in E} \psi_{i,j} (x_i, x_j)
\end{equation}

The goal of inference is to derive the vertices's marginal distributions $P(x_i)$. This is intractable in general, however BP can be used to find exact marginals (for trees) or approximate marginals (for graphs containing loops). This is done through the iterative passing of messages along the edges of the graph. Each edge $(i,j)\in E$ has two messages $m_{i\to j},m_{j\to i}$ being passed along it, indicating each vertex's belief about the other's state. The message $m_{i\to j}$ is a distribution, updated as follows:
\begin{equation} \label{message_update}
    m_{i\to j}(x_j) \propto \sum_{x_i \in A_i} \psi_{i,j}(x_i,x_j)\psi_{i}(x_i)\prod_{k\in \Gamma_i \setminus j} m_{k\to i}(x_i)
\end{equation}
where $\Gamma_i$ indicates the neighbors of $v_i$. Each message is initialized to the uniform distribution and normalized between updates. Messages are iterated until $\epsilon$ convergence, at which point we calculate the beliefs at each vertex:
\begin{equation} \label{belief}
    P(X_i=x_i)\approx b_i(x_i) \propto \psi_i(x_i) \prod_{k\in \Gamma_i} m_{k \to i}(x_i)
\end{equation}

\subsection{Message Scheduling for Belief Propagation} \label{message_scheduling}

BP message schedulings differ by the messages that are updated each iteration. LBP simply updates every edge, every iteration in parallel. That is, all messages are updated using the previous iteration's messages. LBP performance has been examined both empirically~\cite{murphy1999loopy} and theoretically~\cite{mooij2007sufficient}.

Asynchronous approaches enforce sequentialism in message updates, updating each message using the most recent messages. That is, a single message is updated, and that update is immediately used to update other messages. If we assume LBP to be a max-norm contraction, ABP has at least as good convergence rate guarantees as LBP~\cite{elidan2006residual}.

Both \cite{elidan2006residual,pmlr-v5-gonzalez09a} build on ABP using \textit{greedy} update schemes. Residual Belief Propagation (RBP) \cite{elidan2006residual} introduces the \textit{residual}, simply defined as:
\begin{equation}\label{message_residual}
    r(m^t_i) = || f_{BP}(m^t)_i - m^t_i || 
\end{equation}
RBP then selects the next message to update asynchronously based on the highest residual. Intuitively, the program focuses its computational effort to parts of the graph where it moves closer to a converged state.

Residual Splash (RS) \cite{pmlr-v5-gonzalez09a} is an extension of RBP for multi-core parallelization. They extend residuals to vertices, where the residual of a vertex is the maximum residual of incoming messages.
Similar to RBP, vertices are selected greedily, however, in RS, a splash, or BFS search of depth $h$ around the vertex, is performed with updates moving sequentially through the BFS tree. RS demonstrates linear speedup in the number of cores. In this paper we explore LBP, RBP, and RS because of their simplicity and good performance in existing work.

\subsection{Related Work}

BP has been implemented on the GPU for specific BP workloads, including stereo matching \cite{grauer2008gpu,brunton2006belief} and error correcting codes \cite{chandrachoodan2012gpu}. Several works specifically explore memory usage, as the unique architecture of the GPU closely ties memory use and performance. Grauer et al. \cite{grauer2010optimizing} explores using registers, shared memory, and local memory for Belief Propagation and their effect on GPU occupancy for the stereo matching problem. Liang et al.\cite{liang2011hardware} shows a general approach for reducing memory usage for BP by storing only the messages along the edges of partitions of the graphs, allowing messages to be stored in faster shared memory. While we do not explore memory use, our message scheduling work combines naturally with the memory work of both of these approaches.

Several works explore different message schedulings on the GPU for specific BP applications. Yang et al.\cite{yang2006real} filters messages to be updated by removing any messages that have already converged. We employ the same filter as one of the filters in our final RnBP scheduling approach. Xiang et al. \cite{xiang2012real} changes BP on a grid-based stereo problem by using directional updates, that is, messages are updated along dimensions of the grid. Of course, this directional update is specific to grid-based models such as ones used in computer vision. Romero et al.\cite{romero2012sequential} constructed an LDPC code structure in such a fashion that the updates could be partitioned so many could be completed in parallel while still maintaining sequentiality overall. In general, we cannot control the problem as in their case, and creating effective message partitions are problem-specific and non-trivial. Our work takes a general approach that can apply to \textit{any} BP problem, and explore message schedulings that have not yet been explored on the GPU, to the authors's best knowledge.

\section{Frontier-Based Belief Propagation on the GPU} \label{sec:frontierbp}

We present all algorithms examined as realizations of a frontier-based BP framework. In this section, we implement several existing schedulings and benchmark their performance. In the next section, we introduce our own GPU-centric scheduling approach, Randomized Belief Propagation. 

To transfer the schedulings onto the synchronous, many-core architecture of the GPU, we utilize a data-centric, frontier-based parallelization framework \cite{wang2017gunrock,pingali2011tao}. We consider the frontier to be the set of messages selected to be updated synchronously and in parallel each iteration. Message schedulings differ on selection of the frontier, but follow the same general structure presented in Algorithm \ref{alg:frontierbp}.

\begin{algorithm}[H]
\caption{Frontier-Based Belief Propagation} \label{alg:frontierbp}
\begin{algorithmic}[1]
\REQUIRE $pgm, \epsilon$
\ENSURE $marginals$
\STATE $converged \gets False$
\WHILE{!$\textit{converged}$} 
    \STATE $frontier \gets \text{GenerateFrontier}(pgm)$
    \STATE $\text{Update}(frontier, pgm)$
    \STATE $\textit{converged} \gets \text{IsConverged}(pgm, \epsilon)$
\ENDWHILE
\RETURN $\text{Marginals}(pgm)$
\end{algorithmic}
\end{algorithm}

\subsection{Greedy Update Frontier Selection}

We use this frontier-based approach to implement several existing schedulings on the GPU, specifically LBP, RBP, and RS. LBP is simple to implement in this framework: every iteration, all the messages are put in the frontier to be updated. RBP and RS rely on greedily selecting updates based on message residuals. In order to explore the trade-off between parallelism and greedy sequentialism, we will simply adjust the greedy approach to select multiple elements as a frontier per iteration as opposed to a single element. We can consider this to be the selection of the top-$k$ values for update each iteration. Adjusting $k$ allows us to adjust parallelism.

For single-core implementations, the primary data structure employed to perform these greedy updates is a Priority Queue. While concurrent priority queues have been developed, they rely on mutual exclusion, and thus are best suited for asynchronous environments, unlike the GPU. Other work in using GPUs for algorithms with Priority Queue based methods have turned to other approaches, involving sort-and-select, binning, or problem division \cite{davidson2014work, park2010gpu}. Several algorithms for direct top-$k$ GPU selection exist \cite{alabi2012fast,monroe2011randomized}, but speedup only occurs for very large problem sizes. We choose to use a simple sort-and-select approach in order to select the top $k$ elements.

\begin{figure}
    \centering
    \includegraphics[scale=0.2]{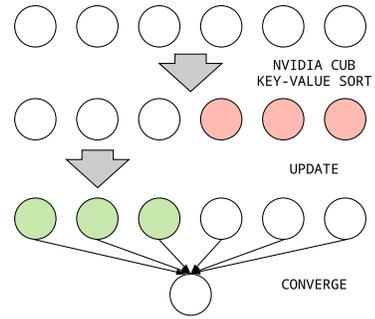}
    \caption{A single bulk-parallel greedy update iteration based on a sort-and-select approach. Each circle represents a message (RBP) or vertex (RS): Green indicates that message/vertex was updated; red indicates that message/vertex has been pruned for this round. Each iteration sorts-and-selects based on the residual, then updates the top-$k$ messages/vertices.}
    \label{fig:bulk_parallel_update}
\end{figure}

We now present the high level approach for our bulk-parallel greedy update selection. We maintain the residuals of either the messages for RBP or the vertices for RS. Each iteration, we perform a key-value sort of the residuals with their corresponding vertices/edges. The top $k$ elements after the sort form the update frontier. RBP updates this frontier directly, RS updates the splash around the selected nodes. A single update is visualized in \figurename \ref{fig:bulk_parallel_update}.

\subsection{Implementation} \label{rbprsimplementation}

We implement LBP, RBP, and RS using Nvidia's CUDA library \cite{nvidia2010programming}. We use a simple adjacency list format for storing graph structure and parameterization. Each edge and vertex is assigned IDs, and for parallel operations, each thread is assigned a subset of the IDs to update. We use the CUDA occupancy API for kernel launch settings and Nvidia's CUB library Radix Sort for the sort operation \cite{merrill2015cuda}. We implement serial RBP (SRBP) as a performance benchmark. We use the same adjacency list format and use the Boost library's Fibonacci heap for the Priority Queue.

\subsection{Benchmarks} \label{benchmarks}

To accurately benchmark performance, we would like to be able to adjust the difficulty of the inference problem. A synthetic benchmark that gives us control over difficulty is the Ising dataset, a standard benchmark for message propagation algorithms \cite{elidan2006residual}. Ising grids are $N\times N$ grids of binary variables. Univariate potentials $\psi_i$ are randomly sampled from the [0,1] range. The pairwise potentials $\psi_{i,j}$ are set to $e^{\lambda C}$ when $x_i = x_j$ and $e^{-\lambda C}$ otherwise. $\lambda$ is sampled from [-0.5,0.5] to make certain potentials favor agreement while others favor disagreement. Varying $C$ changes the difficulty of the inference problem (higher $C$ being more difficult). For RBP and RS, we test on Ising grids of size $100\times 100$ and $200\times 200$, with $C=2.5$. We also run on simpler chain graphs, where $N$ binary variables are formatted in a single long chain. Of course, when a graph is a chain, BP is guaranteed to converge. We sample $\psi_i$ and $\psi_{i,j}$ in the same manner used for our Ising grids. For RBP and RS, we test on chain graphs of size $100000$, with $C=10$.

\subsection{Performance}


\begin{figure*}[!t]
    \centering
    \subfloat[Ising $100\times 100$, $C=2.5$]{
        \includegraphics[scale=0.35]{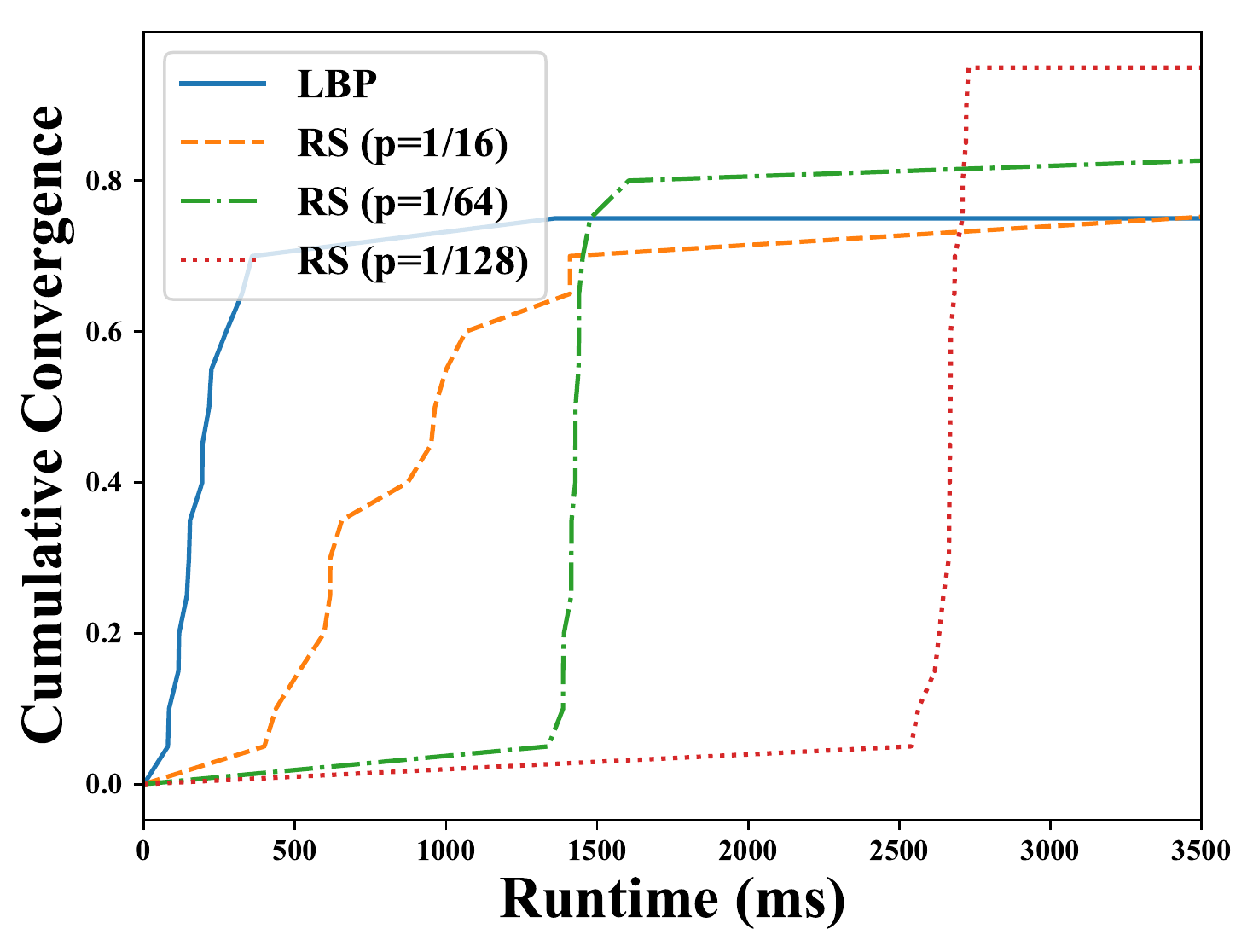}
        \label{rsising100}
    }
    \hfill
    \subfloat[Ising $200\times 200$, $C=2.5$]{
        \includegraphics[scale=0.35]{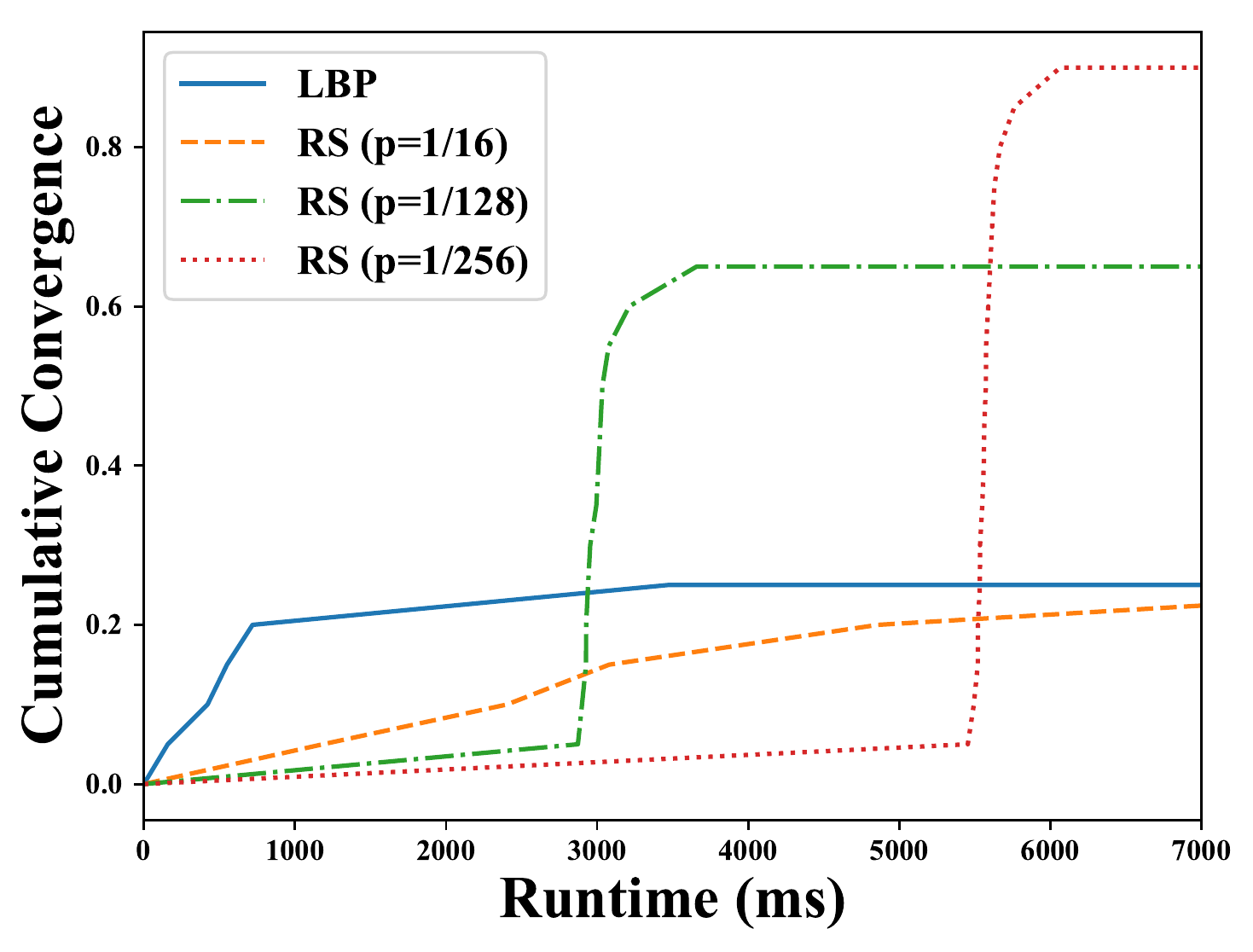}
        \label{rsising200}
    }
    \hfill
    \subfloat[Chain $100000$, $C=10$]{
        \includegraphics[scale=0.35]{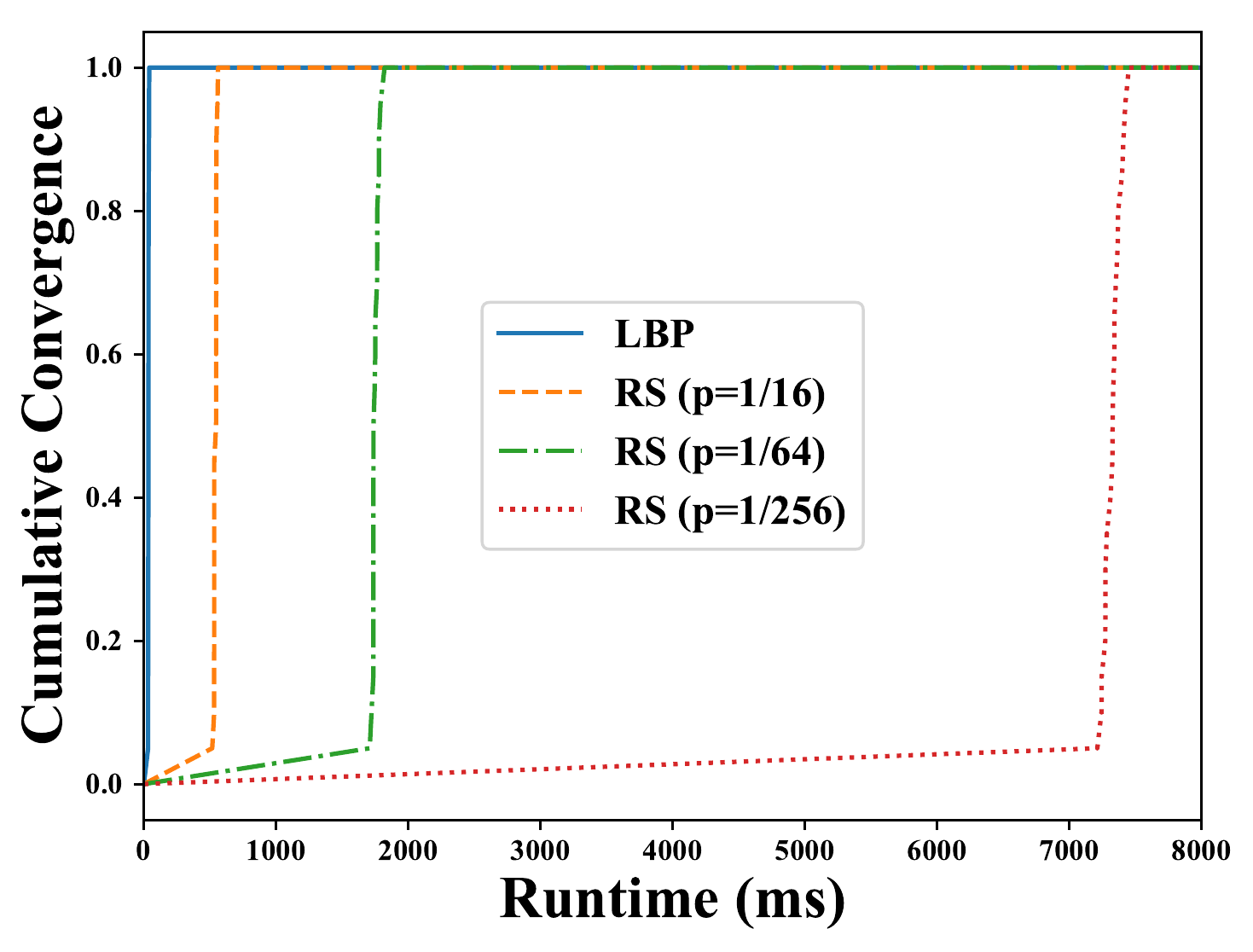}
        \label{rschain}
    }
    \caption{GPU RS cumulative percentage of converged runs for our 2 Ising and 1 Chain dataset, compared to LBP. We see that lower $p$ causes more convergence, at the cost of speed.}
    \label{fig:rsresults}
\end{figure*}

In order to examine parallelism's effect on performance, we introduce a multiplier $p$, where the frontier size each round is $p\cdot 2|E|$. Varying $p$ thus varies the parallelism used. For RS, we lock\footnote{Exploration of different splash depths could be interesting, though we change our focus to randomized updates, and thus do not pursue this further.} splash depth to be $h=2$. We time how long it takes the message updates to converge. Our GPU code is run on a single NVidia Tesla V100 and our CPU code is run on Intel Xeon Processors.

\figurename \ref{fig:rsresults} shows GPU RS performance on our three benchmarks as cumulative convergence graphs, indicating the cumulative percentage of the set of input graphs that have converged as a function of time. GPU RBP exhibits the same patterns on each dataset and thus is not shown for brevity.

Our results indicate that a tradeoff does indeed exist between parallelism and sequentialism. Specifically, we see that as we decrease $p$, that is, we reduce our parallelism, more graphs converge, but they take longer to do so. Thus, low parallelism encourages convergence, while high parallelism encourages speed. LBP, with full parallelism, demonstrates only partial convergence, while RS is able to extend convergence, given time, by reducing parallelism (\figurename \ref{rsising100},\ref{rsising200}).


In Tables \ref{tab:rbpspeedup} and \ref{tab:rsspeedup}, we show the speedup results comparing GPU RBP and RS to SRBP. We compare with the fastest setting in our test runs that converges on all or most of the graphs, indicated for each dataset. For cases where SRBP convergence did not occur (i.e., SRBP failed to converge on all but the Ising $100\times 100$, $C=2.5$ dataset), we provide a conservative lower-bound on speedup based on how long we gave SRBP to run (90 seconds). We see that RS outperforms RBP and both outperform SRBP.

\begin{table}[!t]
    \centering
    \caption{GPU RBP Speedups over SRBP}
    \begin{tabular}{|c|c|c|}
        \hline
        Dataset & Settings & SRBP Speedup \\
        \hline
        Ising $100\times 100$, $C=2.5$ & $p=1/256$ & 3.47x  \\
        \hline
        Ising $200\times 200$, $C=2.5$ & $p=1/256$ & $>4.54$x \\
        \hline
        Chain $100000$, $C=10$ & $p=1/16$ & $>72.31$x \\
        \hline
    \end{tabular}
    \label{tab:rbpspeedup}
\end{table}

\begin{table}[!t]
    \centering
    \caption{GPU RS Speedups over SRBP}
    \begin{tabular}{|c|c|c|}
        \hline
        Dataset & Settings & SRBP Speedup \\
        \hline
        Ising $100\times 100$, $C=2.5$ & $p=1/128$ & 25.85x  \\
        \hline
        Ising $200\times 200$, $C=2.5$ & $p=1/256$ & $>16.08$x \\
        \hline
        Chain $100000$, $C=10$ & $p=1/16$ & $>166.51$x \\
        \hline
    \end{tabular}
    \label{tab:rsspeedup}
\end{table}

There are two primary shortcomings to RBP and RS. First, performance relies heavily upon $p$, and effective $p$ selection is non-trivial. Second, the sort-and-select approach incurs significant overhead. This is best demonstrated by the easy chain dataset (\figurename \ref{rschain}) where RS takes significantly longer than LBP, which converges very quickly. Profiling indicates that on many graphs, both RBP and RS spend more than 90\% of runtime during the sort-and-select step, up to 98\% for certain runs.

\section{Randomized Belief Propagation} \label{sec:rnbp}

To overcome the shortcomings of existing approaches on the GPU and exploit the tradeoff we have demonstrated, we present our novel, low-overhead, randomized message scheduling technique for Belief Propagation on the GPU, Randomized Belief Propagation (RnBP).

\subsection{Algorithm}

We hypothesize that varying the parallelism affects performance more than the specific selection of messages each round when in a many-core environment. We thus perform \textit{random} $k$ selection as opposed to exact top-$k$ selection. 

In creating our message frontier, we employ two filters. In order to encourage selection to be similar to the top-$k$, we only choose the messages to update from those whose residual is above the $\epsilon$ thresholds. Thus, our first filter prunes all messages whose next update will move them less than $\epsilon$.

\begin{figure}
    \centering
    \includegraphics[scale=0.24]{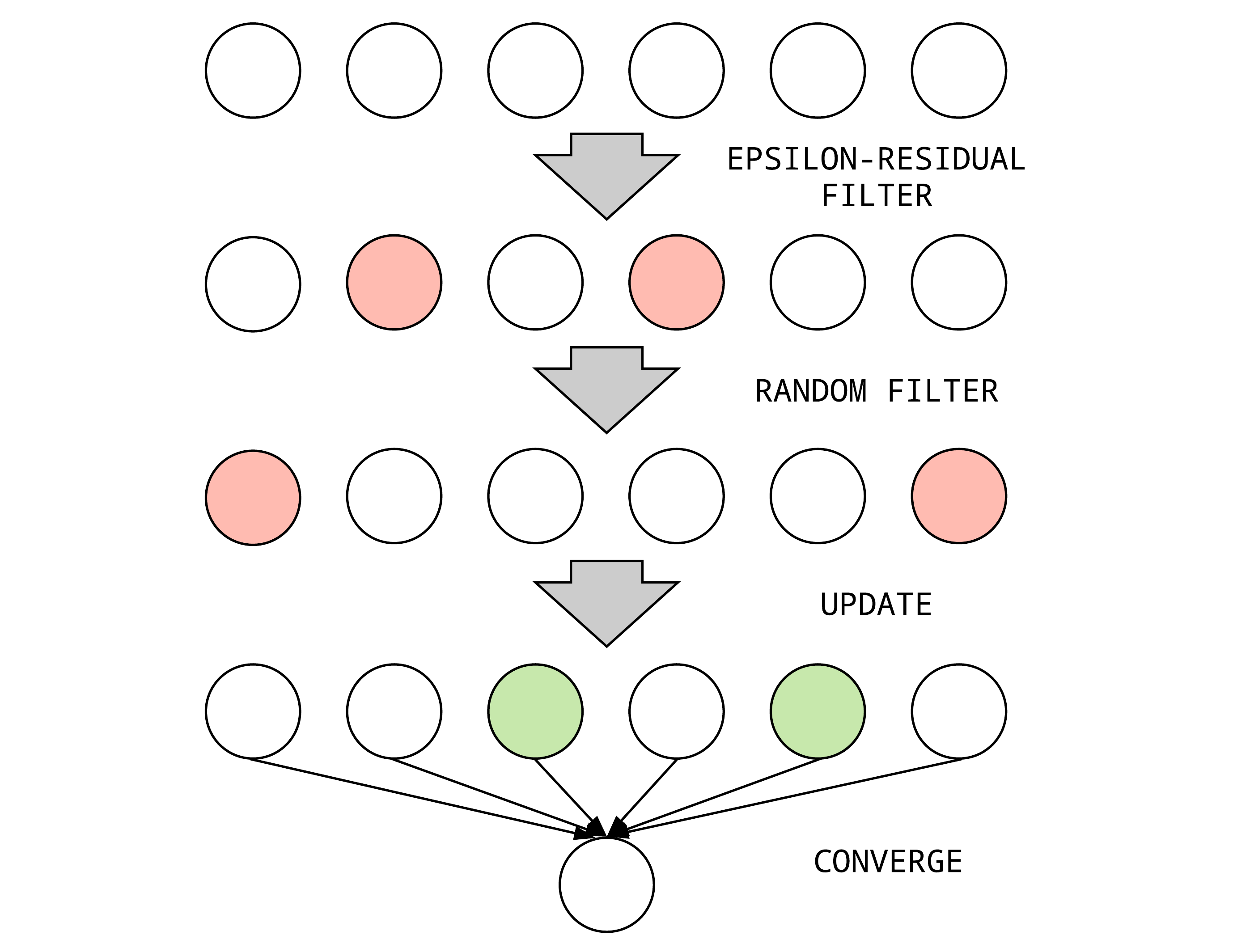}
    \caption{A single RnBP update iteration. Each circle represents a message: green indicates the message was updated; red indicates the message has been pruned for this round. We apply two filters to select the frontier: prune all messages with residual $< \epsilon$ and then we randomly select $p$ of the remaining messages.}
    \label{fig:rnbpupdate}
\end{figure}

The second filter is our randomized filter. We randomly select some percentage $p$ of the remaining messages to be updated. Adjusting $p$ thus allows us to adjust the parallelism for that round. A single update is visualized in \figurename \ref{fig:rnbpupdate}.

Finally, we dynamically range $p$ based on the convergence of the run. Throughout the run, we can track how many of the edges have not converged. The ratio between the number of edges not converged between each iteration becomes an indicator of runtime convergence performance: $EdgeRatio = NewEdgeCount / OldEdgeCount$. If $EdgeRatio$ is low, it is indicative of good convergence, if $EdgeRatio$ is high, it is indicative of bad convergence. We introduce two $p$ settings, one high and one low. We know from our results in Section \ref{sec:frontierbp} that low parallelism encourages convergence and high parallelism encourages speed. Thus, if $EdgeRatio>0.9$, we use the lower parallelism setting, thus encouraging \textit{convergence}. Otherwise, we use the higher parallelism setting, thus encouraging \textit{speed}. We note, overhead prevented similar dynamic $p$ selection from aiding GPU RBP/RS.

\subsection{Implementation}
We implement RnBP in CUDA, using the same data structures, occupancy approach, and parallel operation strategy as described in Section \ref{rbprsimplementation}. For the counting of messages above $\epsilon$, we use Nvidia's CUB library reductions \cite{merrill2015cuda}. We use Nvidia's cuRAND for randomized update selection \cite{nvidia2010curand}.

\subsection{Benchmarks}
We use the same chain and Ising grid benchmarks described in \ref{benchmarks}. We test with Ising grids of size $100\times 100$ with $C=2,2.5,3$ and of size $200\times 200$ with $C=2.5$. For chain graphs, we test with size $100000$ with $C=10$.

\subsection{Performance}

Again, our GPU code is run on a single NVidia Tesla V100 and our CPU code is run on Intel Xeon Processors. We continue to compare to LBP and SRBP. 

As for RBP and RS, we can vary our high and low parallelism settings to get different parallelisms during run time. We found that for our synthetic dataset the high parallelism setting mattered less than the low parallelism setting. As such, we locked our high parallelism to be a full update, thus whenever $EdgeRatio<0.9$, we update the full message frontier update. We show performance on all datasets with low parallelism ($LowP$) being set to 0.7, 0.4, and 0.1.

\begin{figure*}[!t]
    \centering
    \subfloat[Ising $100\times 100$, $C=2$]{
        \includegraphics[scale=0.35]{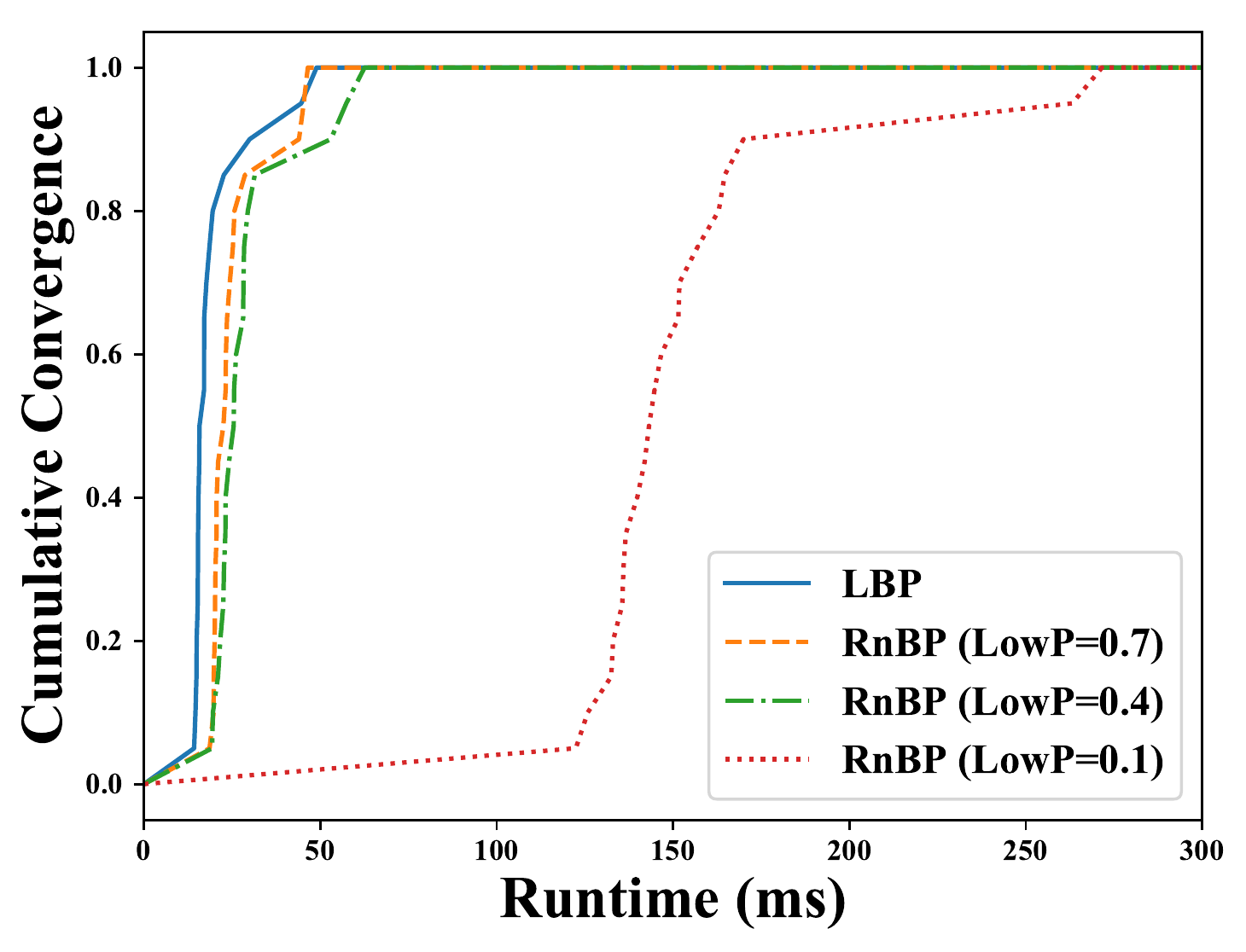}
        \label{rnbpising100c2}
    }
    \hfill
    \subfloat[Ising $100\times 100$, $C=2.5$]{
        \includegraphics[scale=0.35]{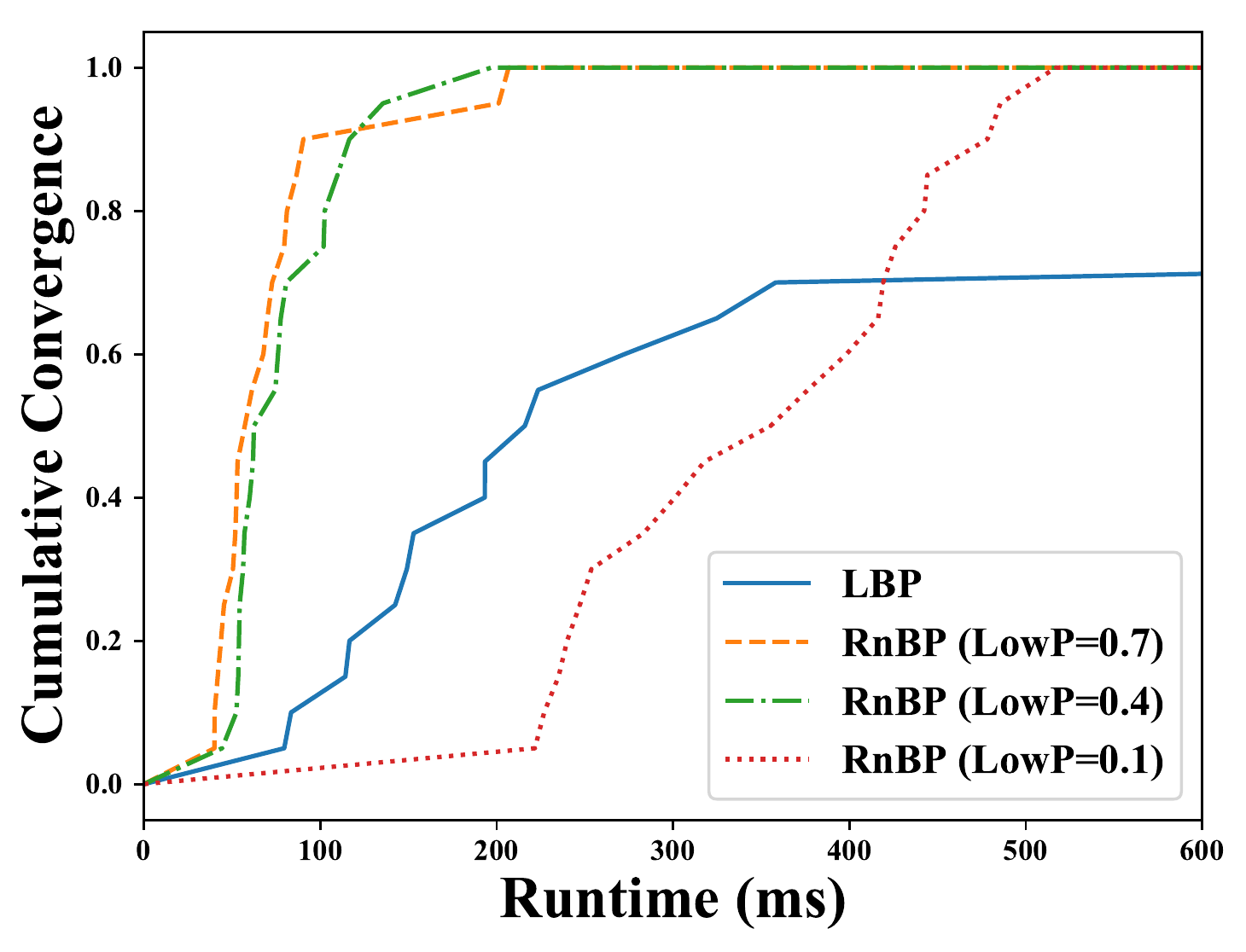}
        \label{rnbpising100c25}
    }
    \hfill
    \subfloat[Ising $100\times 100$, $C=3$]{
        \includegraphics[scale=0.35]{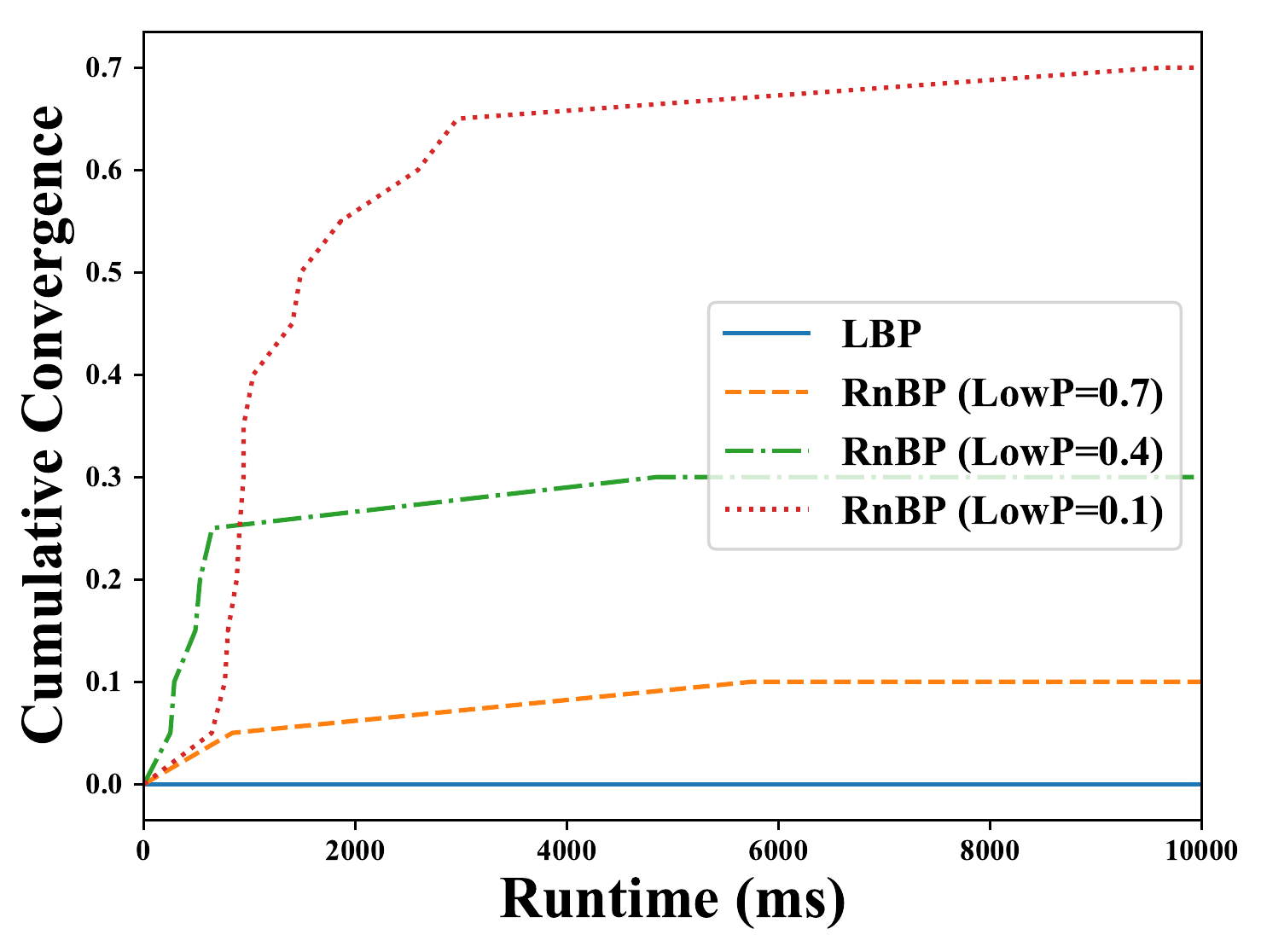}
        \label{rnbpising100c3}
    }
    \hfill
    \subfloat[Chain 100000, $C=10$]{
        \includegraphics[scale=0.35]{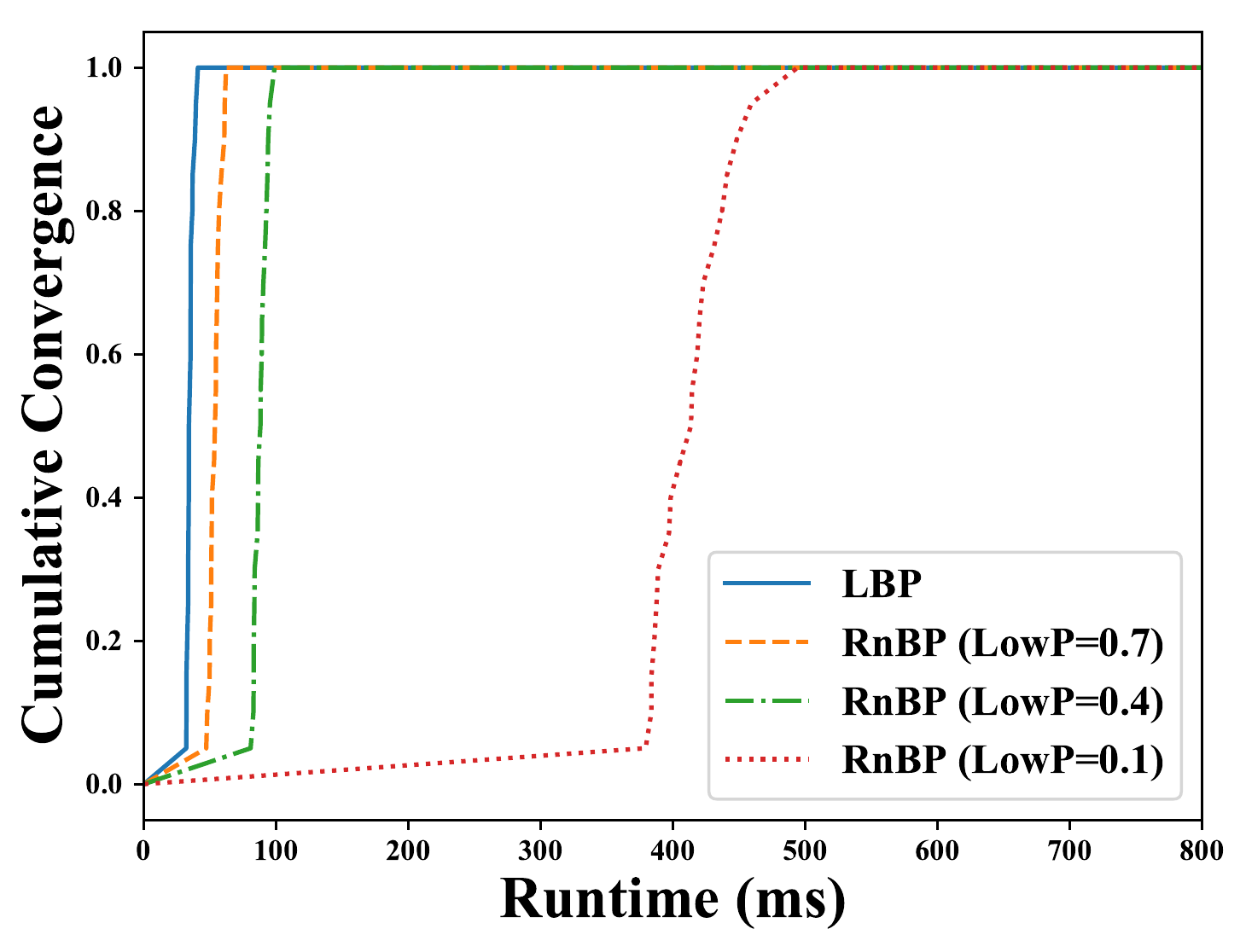}
        \label{rnbpchain100000c10}
    }
    \hfill
    \subfloat[Ising $200\times 200$, $C=2.5$]{
        \includegraphics[scale=0.35]{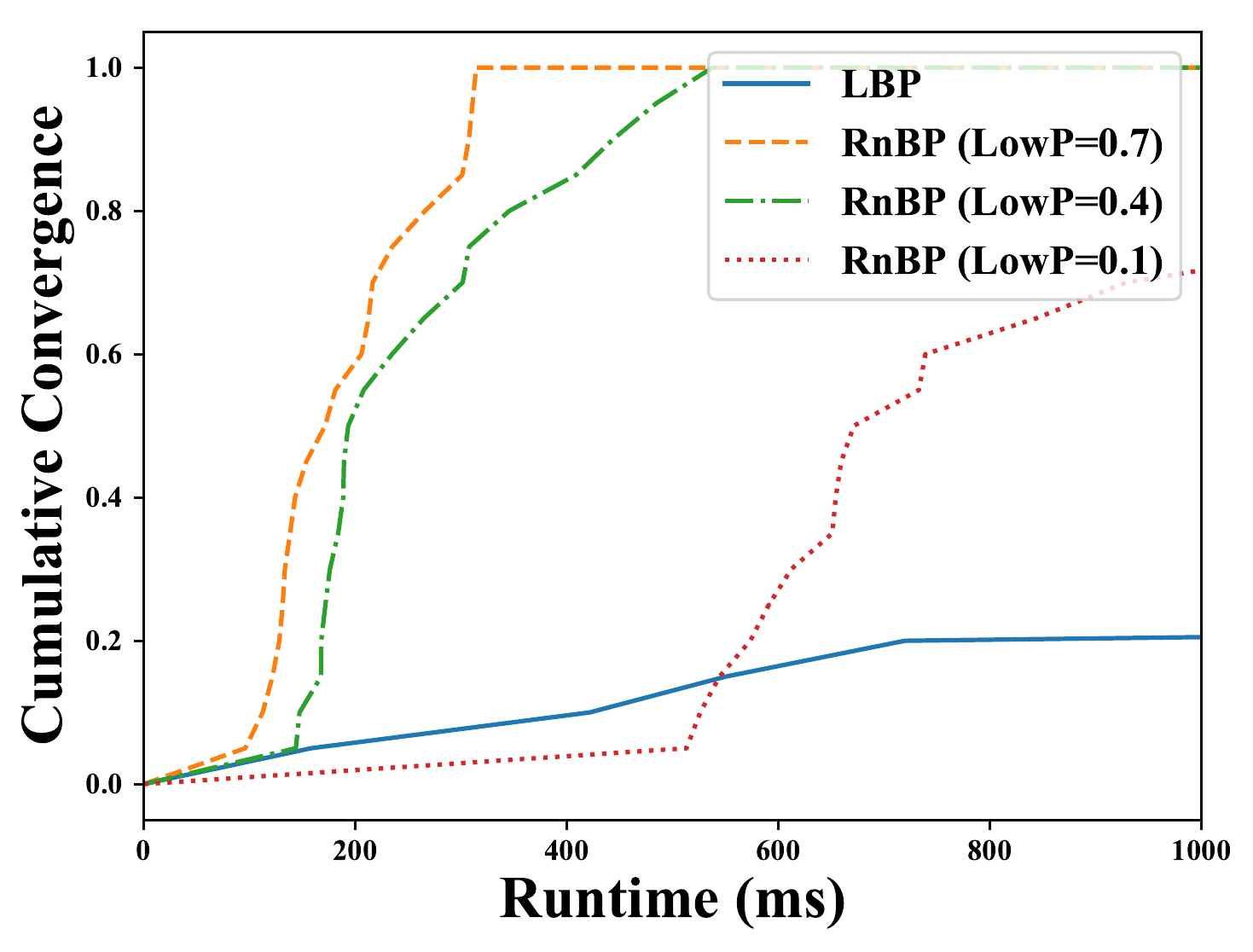}
        \label{rnbpising200c25}
    }
    \hfill
    \subfloat[Protein-Folding Dataset \cite{yanover2003approximate}]{
        \includegraphics[scale=0.35]{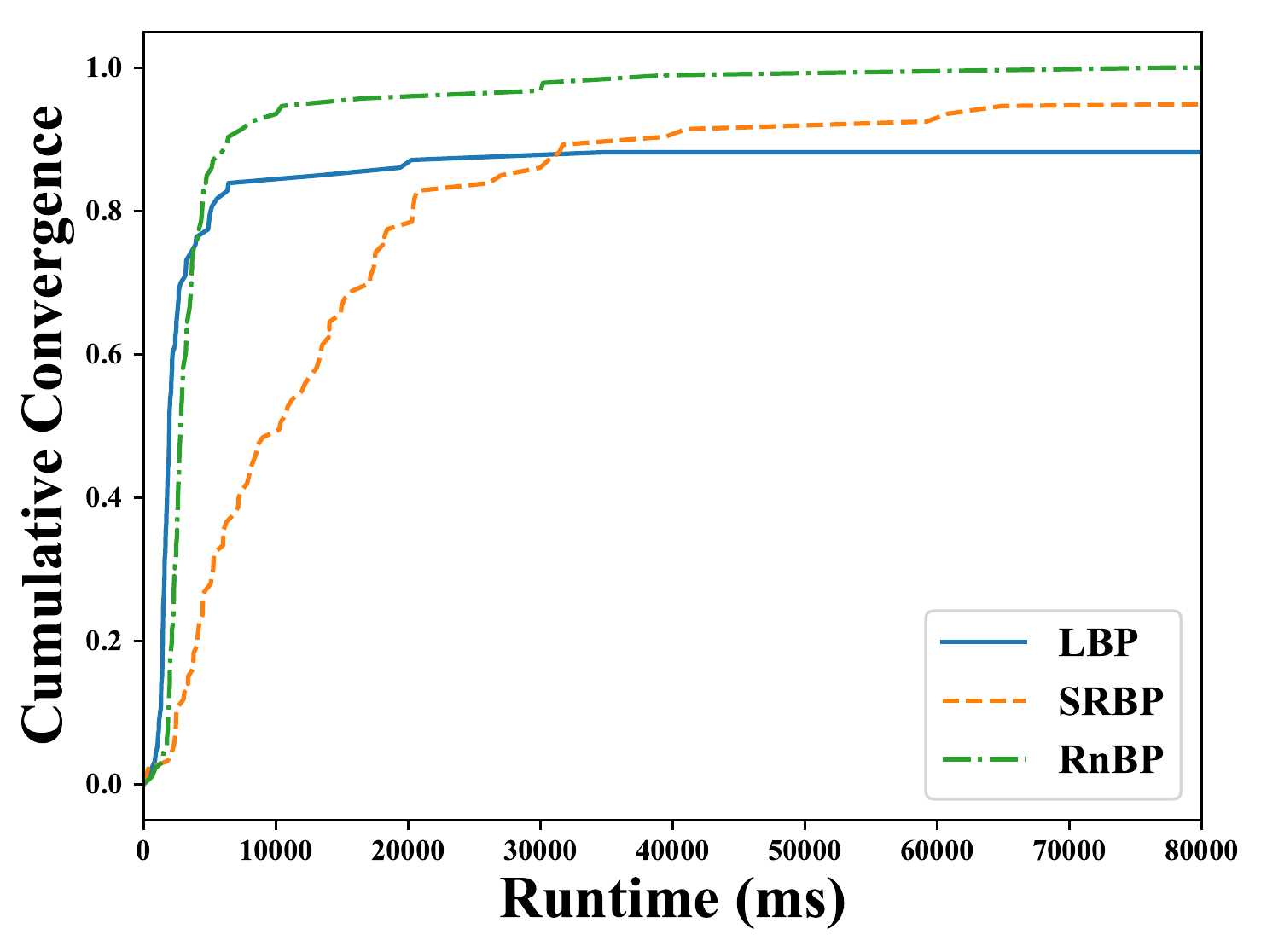}
        \label{fig:protein-perf}
    }
    \caption{GPU RnBP cumulative percentage of converged runs for 5 Ising, 1 Chain, and 1 Protein-Folding datasets, compared to GPU LBP.}
    \label{fig:rnbpresults}
\end{figure*}

\figurename \ref{rnbpising100c2}-\ref{rnbpising200c25} shows GPU RnBP performance on our benchmarks as cumulative convergence graphs. For easy graph datasets, where LBP converges for most or all, we notice that RnBP with higher parallelism settings (i.e., $LowP=0.7,0.4$) nearly matches LBP performance (see \figurename \ref{rnbpising100c2},\ref{rnbpchain100000c10}). This shows the value in RnBP's lack of overhead. As the graphs become more difficult, where LBP only converges on some, we see that RnBP continues to converge quickly on \textit{all} graphs (see \figurename \ref{rnbpising100c25}, \ref{rnbpising200c25}). RnBP converges with much higher parallelism than that required for RS and RBP. Using the higher parallelism settings allows speed paired with convergence. We see that this allows for RnBP to actually provide speedups \textit{over} GPU LBP runtimes (\figurename \ref{rnbpising100c25},\ref{rnbpising200c25}), averaging 9x speedups on the Ising 200$\times$200, C=2.5 dataset.

Notice, LBP fails to converge on any graphs for the difficult 100$\times$100, C=3 dataset. We see that we can effectively drop parallelism in RnBP, however, to encourage convergence (see \figurename \ref{rnbpising100c3}). We do so without significant overheads yielding dramatic slow downs. This convergence behavior applies to larger and more difficult graphs than the ones RBP and RS could handle. RnBP thus \textit{extends the classes of Belief Propagation problems for which GPU speedups can be applied}. We note that for the difficult dataset, RnBP can still be sensitive to the selected parallelism. However, on all our other datasets, RnBP is fairly robust to parallelism selection. Thus, while not completely solved, RnBP is a considerable improvement to existing approaches.

We characterize the speedup of RnBP over SRBP in Table \ref{tab:rnbpsrbp}. Again, we compare with the fastest setting in our test runs that converges on all or most of the graphs, indicated for each dataset, and present conservative lower bounds when SRBP failed to converge (given 90 seconds).

\begin{table}[!t]
    \centering
    \caption{GPU RnBP Speedups over SRBP}
    \begin{tabular}{|c|c|c|}
        \hline
        Dataset & Settings & SRBP Speedup \\
        \hline
        Ising $100\times 100$, $C=2$ & $LowP=0.7$ & 2203.58x  \\
        \hline
        Ising $100\times 100$, $C=2.5$ & $LowP=0.7$ & 1135.05x  \\
        \hline
        Ising $100\times 100$, $C=3$ & $LowP=0.1$ & 61.28x  \\
        \hline
        Ising $200\times 200$, $C=2.5$ & $LowP=0.7$ & $>529.997$x \\
        \hline
        Chain $100000$, $C=10$ & $LowP=0.7$ & $>1676.92$x \\
        \hline
    \end{tabular}
    \label{tab:rnbpsrbp}
\end{table}


\subsection{Additional Tests}

\begin{figure}
    \centering
    \includegraphics[scale=0.4]{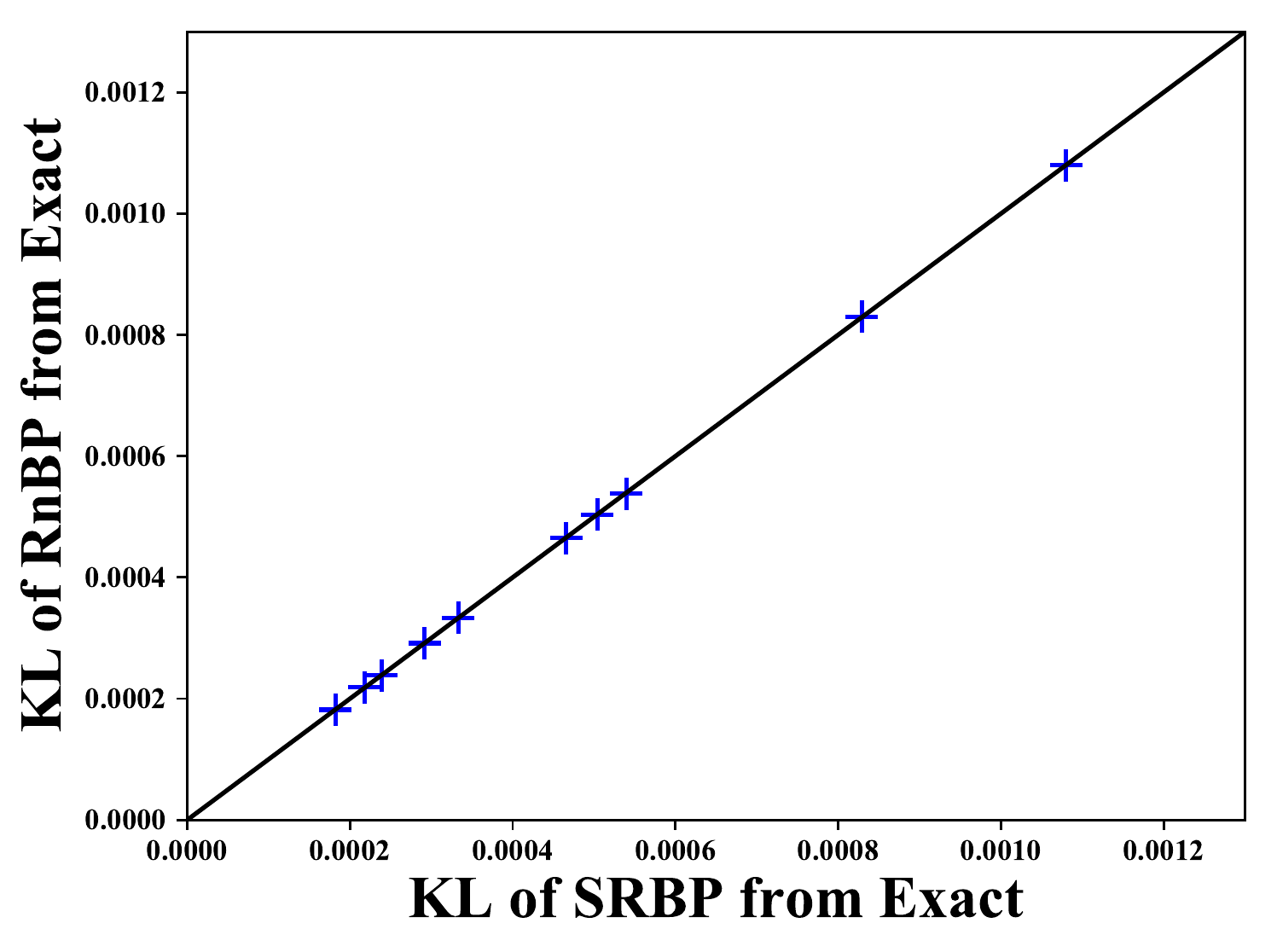}
    \caption{Correctness of the converged marginals for GPU RnBP and SRBP as compared to the exact marginals for Ising $10\times 10$, $C=2$ dataset.}
    \label{fig:correctness_graph}
\end{figure}

As RnBP is a novel message scheduling, we provide several additional tests to examine performance. To test correctness, we created a smaller Ising dataset, size $10\times 10$, $C=2$, for which exact inference is tractable. We use Variable Elimination to find the exact marginal values, then determine the KL-divergence between the exact results and the results of both SRBP and RnBP (run with $LowP=0.7$). These are shown in figure \ref{fig:correctness_graph}. We see that RnBP achieves the same quality of result as compared to SRBP.

We tested RnBP on a real-world dataset, specifically a protein-folding dataset \cite{yanover2003approximate}. This dataset contains graphs with vertices representing amino acid units and the setting at each vertex representing the side-chain configuration. The possible settings at each vertex ranges from 2 to 81 and the graph structure is highly irregular. The cumulative convergence is shown in \figurename \ref{fig:protein-perf} (We run RnBP with $LowP=0.4, HighP=0.9$). Despite the different structure as compared to our synthetic dataset and without any finetuning to handle load-imbalanced message updates, we see that RnBP yields fast, convergent performance. Given 3 minutes per graph, RnBP was the only approach to converge on all graphs and yielded an average of 4.4x speedup over SRBP when SRBP converged.

\section{Conclusions and Future Work} \label{sec:conclusion}

In this work, we presented a study of message scheduling approaches for BP on many-core GPU systems (summarized in Table \ref{tab:approaches}). We hypothesized the existence of a tradeoff between parallelism and sequentialism for BP speed and convergence, and that GPUs could be used to exploit that tradeoff for performant BP. We presented many-core, frontier-based implementations for two asynchronous message schedulings, RBP \cite{elidan2006residual} and RS \cite{pmlr-v5-gonzalez09a}, and showed empirically that indeed a tradeoff exists. Specifically, lower parallelism encourages convergence, while higher parallelism encourages speed. We also show that these approaches incur significant overhead, suggesting that a new GPU-centric approach is needed. In this direction, we presented a novel message scheduling we call Randomized Belief Propagation (RnBP), which utilizes randomization to select frontiers for updating. We demonstrate that this approach yields higher convergence while maintaining speed, providing speedups over serial and existing GPU methods on both synthetic and real-world datasets. Our implementation is available online\footnote{https://github.com/mvandermerwe/BP-GPU-Message-Scheduling}.

\begin{table}[!t]
    \centering
    \caption{Algorithms Explored (Bold indicates contribution)}
    \begin{tabular}{|c||c|c|}
        \hline
        Algorithm & Frontier Selection & Many-Core \\
        \hline
        GPU LBP & All Messages & \checkmark \\
        Serial RBP/RS & Priority Queue & \text{\sffamily X} \\
        \textbf{GPU RBP/RS} & Sort-and-Select & \checkmark \\
        \textbf{GPU RnBP} & Randomized & \checkmark \\
        \hline
    \end{tabular}
    \label{tab:approaches}
\end{table}

Our approach is a general solution for BP, meaning it can be integrated naturally with many variants of BP \cite{felzenszwalb2006efficient, yedidia2001generalized} as well as with GPU memory improvements \cite{grauer2010optimizing,liang2011hardware}, to further extend performance.

We think that this work could be of interest to other algorithms that utilize iterative, convergent graph updates, such as the Generalized Distributive Law \cite{aji2000generalized}, of which BP is a subclass, and  Graph Neural Networks \cite{liao2018graph}.

\section*{Acknowledgment}
This work was supported in part by NSF awards 1704715 and 1817073.

\bibliographystyle{IEEEtran}
\bibliography{IEEEabrv,bibliography}

\end{document}